\documentclass[aps,pra,twocolumn,superscriptaddress]{revtex4-1}

\usepackage[english]{babel}
\usepackage{braket}
\usepackage{mathtools}
\usepackage{tikz}
\usepackage{tikz-3dplot}
\usepackage{amssymb,bm}
\usepackage{tabularx, booktabs}
\usepackage{tikz,pgfplots}
\usepackage{amsmath,amsfonts,mathrsfs,amssymb}
\usepackage{url}
\usepackage{graphicx}
\usepackage{epsfig}
\usepackage{color}
\usepackage{array}
\usepackage{subcaption}
\usepackage{csquotes}
\usepackage{xpatch}
\usepackage{xspace}
\usepackage[normalem]{ulem} 
\usepackage{physics}
\usepackage{appendix}
\usepackage{bbold} 

\usepackage[colorlinks=true, linkcolor=blue, citecolor=dred]{hyperref}
\hypersetup{
    colorlinks=true,
    citecolor=gray,
    filecolor=blue,
    linkcolor=blue,
    urlcolor=red
}

\definecolor{dgreen}{rgb}{0,.5,0}
\definecolor{dred}{rgb}{.7,.0,.0}
\definecolor{ICGMmarine}{rgb}{0.168, 0.168, 0.525}
\definecolor{ICGMblue}{rgb}{0, 0.549, 0.714}
\definecolor{ICGMorange}{rgb}{0.968, 0.647, 0}
\definecolor{ICGMyellow}{rgb}{1, 0.804, 0}

\newcommand{\Eq}[1]{Eq.~(\ref{#1})}

\newcommand{\Fig}[1]{Fig. \ref{#1}}

\newcommand{\Sec}[1]{Sec. \ref{#1}}

\newcommand{\etal}{{\it et al.}}

\newcommand{\be}{\begin{eqnarray}}
\newcommand{\ee}{\end{eqnarray}}

\tikzset{>=latex} 
\tikzstyle{proj}=[projcol!80,line width=0.08] 
\tikzstyle{area}=[draw=ICGMmarine,fill=veccol!80,fill opacity=0.6]
\tikzstyle{vector}=[-stealth,ICGMmarine,thick,line cap=round]
\tikzstyle{unit vector}=[->,ICGMmarine,thick,line cap=round]
\tikzstyle{dark unit vector}=[unit vector,ICGMmarine]
\usetikzlibrary{angles,quotes} 
\usetikzlibrary{shapes,shadows}
\tikzstyle{startstop} = [rectangle, rounded corners, minimum width=3cm, minimum height=1cm, inner sep=3pt, text centered, draw=black, fill=ICGMorange!30]
\tikzstyle{process} = [rectangle, minimum width=3cm, minimum height=1cm, inner sep=3pt, text centered, draw=black, fill=ICGMblue!30]
\tikzstyle{arrow} = [thick,->,>=stealth]
\usetikzlibrary{arrows.meta}
\usepackage{xr}

\begin{document}

\title{A versatile unitary transformation framework for an optimal bath construction in density-matrix based quantum embedding approaches}

\author{Quentin Mar\'ecat}
\email{quentin.marécat@umontpellier.fr}
\affiliation{ICGM, Université de Montpellier, CNRS, ENSCM, 34000 Montpellier (France)}
\author{Matthieu Sauban\`ere}
\affiliation{ICGM, Université de Montpellier, CNRS, ENSCM, 34000 Montpellier (France)}

\begin{abstract}
The performance of embedding methods is directly tied to the quality of the bath orbitals construction. In this paper, we develop a versatile framework, enabling the investigation of the optimal construction of the orbitals of the bath.
 As of today, in state-of-the-art embedding methods, the orbitals of the bath are constructed by performing a Singular Value Decomposition (SVD) on the impurity-environment part of the 1RDM, as originally presented in Density Matrix Embedding Theory (DMET). Recently, the equivalence between the SVD protocol and the use of unitary transformation, the so-called Block-Householder transformation, has been established. 
We present a generalization of the Block-Householder transformation by introducing additional flexible parameters. The additional parameters are optimized such that the bath-orbitals fulfill physically motivated constrains. The efficiency of the approach is discussed and examplified in the context of the half-filled Hubbard model in one-dimension.
\end{abstract}
\maketitle
\section{Introduction}
Quantum chemistry is built on the foundation of solving the Schrödinger equation, an endeavor that quickly becomes computationally prohibitive for systems of practical interest. The challenge comes from the exponentially scaling cost with system size.
In that context, quantum embedding methods~\cite{sun_quantum_2016,wasserman_quantum_2020} have emerged as a powerful tool, particularly for the study of strongly correlated systems~\cite{kotliar_cellular_2001,zheng_self-consistent_1993,ma_quantum_2021}, where traditional methods often fall short~\cite{cohen_insights_2008,burke2012perspective}. 
In short, an embedding protocol consists in partitioning the original extended system into fragments. Each fragment, often referred to as an \textit{impurity}, is complemented with bath orbitals to define an effective reduced system described by an effective Hamiltonian. The reduced system, comprising the fragment+bath (denoted the \textit{cluster}), contains $N_i$ orbitals of the fragment, which interact with the additional $N_b$ bath orbitals. Ideally, the fragment and bath reduced system is entirely decoupled from the rest of the system, which is referred to as the \textit{environment} of the cluster.
The cornerstone of these embedding methods is the construction of bath orbitals associated with a fragment in order to derive the effective Hamiltonian of the reduced system. This effective Hamiltonian captures the physics of the larger original environment while integrating out most of its degrees of freedom. Consequently, it provides a highly efficient approach to treat localized electron correlation without the need to explicitly consider the entire system, thus balancing computational feasibility with accuracy. 
Different embedding strategies have been proposed in the literature to construct such a reduced effective Hamiltonian.

Among these embedding strategies, some rely on the Green's function formalism, which focuses on single-particle excitations and can naturally incorporate electronic correlation effects. In particular, within the dynamical mean-field theory (DMFT) formalism~\cite{muller-hartmann_correlated_1988,georges_dynamical_1996,georges_hubbard_1992}, the effective Hamiltonian is an effective Anderson impurity model (AIM)~\cite{anderson1961localized}, and is derived self-consistently using the Green's function of the system. Note that various alternative embedding approaches using the Green's function have been developed to derive effective AIM~\cite{potthoff2003self,sarker1988new,mazouin2019site} or other effective Hamiltonians~\cite{nguyen_lan_rigorous_2016,lupo_maximally_2021}.

In parallel, embedding approaches have been proposed within the one-body Reduced Density-Matrix formalism (1RDM).
The aim is to design a protocol to construct the bath orbitals, and the corresponding effective Hamiltonian of the cluster, that become functional of the 1RDM. 
Among the different approaches recently proposed~\cite{lanata2023derivation,senjean_projected_2019,sekaran2022local,mitra2023density}, the pioneering work of Knizia \textit{et al.}~\cite{knizia_density_2012} proposes to define the effective Hamiltonians by means of the Schmidt decomposition of a single Slater Determinant (SD)  $|\Phi\rangle$ that is univocally associated with an idempotent 1RDM $\gamma^0$, such that the Schmidt decomposition of $|\Phi\rangle$ defines a projector via a Singular Value Decomposition (SVD) of the fragment-environment part of $\gamma^0$~\cite{wouters_practical_2016,yalouz_quantum_2022,sekaran_unified_2023}. Indeed, SVD is known to yield the most compact representation of all informations contained within the fragment-environment 1RDM.
This approach imposes that the bath constructed through the SVD contains as many orbitals as the impurity fragment, and also that only idempotent 1RDM can be used to describe the extended system. Over the past decade, different variations and improvements of the original DMET have been investigated and benchmarked~\cite{ayral_dynamical_2017,cances2023mathematical,sun_finite_2020,ye_incremental_2018,hermes_multi_2019}.
Recently, Sekaran {\it et al.}~\cite{sekaran_householder_2021,yalouz_quantum_2022} proposed to use a specific unitary transformation, defined as a functional of the 1RDM, to construct the bath orbitals~\cite{sekaran_householder_2021}. The aforementioned unitary transformation is known as the Block-Householder transformation. They demonstrated that the SVD of the fragment-environment 1RDM is equivalent to the Block-Householder transformation, even for a non-idempotent 1RDM~\cite{sekaran_unified_2023}. 
Despite the appealing compactness of the resulting sub-space of the bath, it should be noted that there is no inherent physical reason to believe that such decomposition yields the "best" effective Hamiltonian for reproducing the interactions between the fragment and its environment.

In this contribution, we propose a versatile framework that generalizes the block-Householder transformation and introduces additional degrees of freedom for the construction of the orbitals of the bath and the derivation of the effective Hamiltonian. As a result, additional constraints are required for the bath orbitals such as maximally disentangling the cluster from the environment, or matching density matrices. The effects of the additional constraints to optimally construct the orbitals of the bath are benchmarked on the well-known but non-trivial half-band filled one-dimensional Hubbard model~\cite{hubbard_electron_1963}, following the divide and conquer algorithm proposed in a previous work by the authors~\cite{marecat2023unitary}.

\section{Theory}
Let us consider the paramagnetic Hubbard ring given by
  \begin{eqnarray}
  \label{eq:hubbard}
    \hat{H} =    -t\sum_{<ij>\sigma} \hat{c}^\dagger_{i\sigma}\hat{c}_{j\sigma}, +  U\sum_i \hat{n}_{i\uparrow} \hat{n}_{i\downarrow}
  \end{eqnarray}
  where $\hat{c}^\dagger_{i\sigma}$ ($\hat{c}_{i\sigma}$) corresponds to the creation (annihilation) of an electron of spin $\sigma$ on the $i$-th orbital, $\hat{n}_{i\sigma}$ is the counting operator equal to $\hat{c}^\dagger_{i\sigma}c_{i\sigma}$. Indexes $<ij>$ refer at nearest neighbor orbitals. $-t$ corresponds to the hopping integral, while $U$ stands for the Coulomb integral.
  
  \subsection{Quantum bath from the Block-Householder transformation}\label{sec:review}

In this section we recall the generic construction of the unitary Block-Householder transformation $\mathbf{R}^\sigma$, following the work of Sekaran \etal~\cite{sekaran_householder_2021}.
The unitary transformation $\mathbf{R}^\sigma$ is defined as a functional of the spin 1RDM $\gamma^\sigma$. It performs the rotation of the mono-electronic basis set such that  the system is divided into a compact subset of $N_i + N_b$ orbitals, the cluster, interacting as few as possible with a large number of $N_e$ orbitals.
 More precisely, the unitary transformation is design such that {\it (i)} the fragment in the full system is the same as in the cluster\\
  \begin{eqnarray} \label{eq:iden}
&\tilde{c}_{i\sigma}^{\dagger} =\hat{c}_{i\sigma}^{\dagger},\quad \tilde{c}_{i\sigma} = \hat{c}_{i\sigma} \quad \forall i \in \rm fragment,  
\end{eqnarray}
where
\begin{eqnarray} \label{eq:transfo1}
   \tilde{c}_{i\sigma}^{\dagger} = \sum_{k}\mathbf{R}^{\sigma \dagger}_{ik}\hat{c}_{k\sigma}^{\dagger}, &\quad \tilde{c}_{i\sigma} = \sum_{k}\mathbf{R}^{\sigma}_{ik}\hat{c}_{k\sigma},
\end{eqnarray}
stand for the creation and annihilation operators of an electron in the $k$-th orbital with spin $\sigma$ expressed in the $\mathbf{R}^\sigma$ representation,
 and {\it (ii)} the fragment is fully disconnected from the environment at the one-body level,
 \begin{eqnarray} 
 \label{eq:uncoupl}
   &\sum_k \mathbf{R}^\sigma_{ik} \gamma^\sigma_{ik} = 0 \quad  \forall i \in \text{ fragment},  \; \forall l \notin \rm cluster.
\end{eqnarray}
Numerous unitary transformations satisfy conditions (\ref{eq:iden}) and (\ref{eq:uncoupl}) resulting in identical bath orbitals subspace~\cite{tows2011lattice,schade2018adaptive}.
Among all these transformations, we focus explicitly on the Block-Householder transformation  $\mathbf{R}^\sigma$  defined using an auxiliary matrix $\mathbf{V}[\gamma^\sigma] \in \mathbb{R}_{NN_i}$, with 
\begin{eqnarray}
\mathbf{R}^\sigma = Id. - 2\mathbf{V}\left(\mathbf{V}^{ T}\mathbf{V}^{} \right)^{-1}\mathbf{V}^{T}.
\end{eqnarray}
By construction, $\mathbf{R}^\sigma$ is a normal involution (i.e. $\mathbf{R}^{\sigma^{-1}}$=$\mathbf{R}^\sigma$, $\mathbf{R}^{\sigma^T}=\mathbf{R}^\sigma$ and det$(\mathbf{R}^\sigma)=(-1)^{N_i}$) with eigenvalues $\{1,-1\}$, and dim(Ker($\mathbf{R}^\sigma + Id.$))$=r=N_i$. 
The matrix $\mathbf{V}$ is given as following\\
\begin{equation}
  \label{eq:V_Blk_House}
\mathbf{V} = \left(
\begin{array}{l}
\mathbf{0}^r_{N_i}\\
\gamma^{r}_{N_i:2N_i} + \mathbf{X}^{r}_{N_i}\\
\gamma^{r}_{2N_i:N}
\end{array}
\right).
\end{equation}
The superscript $\; ^{r}\;$ (underscript $\; _{N_i}\;$) stands for th number of columns (lines) of the associated matrix, and the notation $\mathbf{A}^{c}_{i:j}$  corresponds to consider only the elements from $i$-th to the $j$-th lines for first $c$-th columns of the matrix $\mathbf{A}$. 
For example, the square matrix $\gamma^{r}_{N_i:2N_i}$ refers to the  first $r=N_i$-th columns of $\gamma^\sigma$ from the $N_i$-th element to the $2N_i$-th.
The null matrix $\mathbf{0}^r_{N_i}$ allows to fix the identity in equation~(\ref{eq:iden}) on the subspace corresponding to the fragment, and 
$\mathbf{X}^{r}_{N_i}$ is a square matrix to be determined  in order to satisfy condition~(\ref{eq:uncoupl}). 
To that aim, Rotella \etal~\cite{rotella_applied_nodate} proposed a systematic construction with \\
 \begin{eqnarray} \label{eq:X}
\mathbf{X}^{r}_{N_i} = \mathbf{P}^T \sqrt{\mathbf{D}} \mathbf{P}\gamma^r_{N_i} ,
 \end{eqnarray}
and $\sqrt{\mathbf{D}}={\rm diag}^N_{i=1}\{\sqrt{d_i}\}$, where the nonnegative scalar $d_i$ and the orthogonal matrix $\mathbf{P}$ are defined by
 \begin{eqnarray}\label{eq:X_diag}
Id. + \left(\gamma^r_{2N_i:N}\gamma^{r^{-1}}_{N_i:2N_i}\right)^T\left(\gamma^r_{2N_i:N}\gamma^{r^{-1}}_{N_i:2N_i} \right) = \mathbf{P}^T\mathbf{D}\mathbf{P},
 \end{eqnarray}
 and $\mathbf{D}$ ($\mathbf{P}$) is eigenvalues (eigenvectors) matrix of the left sided  matrix in equation~(\ref{eq:X_diag}), respectively. The construction of $ \mathbf{X}^{r}_{N_i}$  in equation~(\ref{eq:X}) holds only if  the square matrix $\gamma^{r}_{N_i:2N_i}$ is invertible.
Then we obtain\\
\begin{equation} \label{eq:Rprop}
\left( \mathbf{R}^\sigma \gamma^\sigma \right)^r = \left(
\begin{array}{l}
\gamma^r_{N_i}\\
-\mathbf{X}^r_{N_i}\\
\mathbf{0}^r_{N_e}
\end{array}
\right) .
\end{equation} 
Equation~(\ref{eq:Rprop}) is strictly equivalent as the conditions defined in equations~(\ref{eq:iden}) and~(\ref{eq:uncoupl}), {\it i.e.} the transformation preserves the orbitals of the fragment in the cluster and disconnects the fragment from the environment at the one-body level.

Recent study expounded on the mathematical equivalence between the bath orbitals constructed using the Block-Householder transformation of the 1RDM and the SVD of the fragment-environment 1RDM~\cite{sekaran_unified_2023}. As a result of this equivalence, the Block-Householder transformation provides the most compact representation, encapsulating information pertaining to the one-body level interactions between the fragment and its environment.
Once the Householder transformation has been presented to construct the orbitals of the bath, we briefly recall the procedure based on $\mathbf{R}^\sigma$ in order to define an effective, yet approximated,  embedded  Hamiltonian. 
Details and discussion on the straightforward generalization to the {\it ab-initio} Hamiltonian can be found in Ref.~\cite{marecat2023unitary}.
Following equation~(\ref{eq:transfo1}), we perform the transformation of the full Hamiltonian~(\ref{eq:hubbard}) to derive the Hamiltonian $\tilde{H}$,
\begin{eqnarray} \label{eq:H-newrep}
\tilde{H} = \sum_{ij\sigma}\tilde{t}_{ij}\tilde{c}^\dagger_{i\sigma}\tilde{c}_{j\sigma} + \sum_{ijkl}\tilde{U}_{ijkl}\tilde{c}^\dagger_{i\uparrow}\tilde{c}_{j\uparrow} \tilde{c}^\dagger_{k\downarrow}\tilde{c}_{l\downarrow},
\end{eqnarray}
where single- and two-bodies integrals in the block-Householder representation $\tilde{t}_{ij}$ and  $\tilde{U}_{ijkl}$, respectively,  are calculated in the same manner,
\begin{align}
\tilde{t}_{ij} &= \sum_{kl} t_{kl}\mathbf{R}^{\sigma\dagger}_{ik}\mathbf{R}^\sigma_{jl},\label{eq:tce} \\
\tilde{U}_{ijkl} &= U\sum_m\mathbf{R}^{\sigma\dagger}_{im}\mathbf{R}^\sigma_{jm}\mathbf{R}^{\bar{\sigma}\dagger}_{km}\mathbf{R}^{\bar{\sigma}}_{lm}.\label{eq:Uijkl}
\end{align}

Considering that $\mathbf{R}^\sigma$  maximally uncouples the fragment from the environment though the orbitals of the bath, an approximation for the effective Hamiltonian on the cluster $\bar{H}^c$ is obtained by projecting the Hamiltonian~(\ref{eq:H-newrep}) onto the cluster orbitals.
Part of the cluster-environment two-body interactions are taken into account at a mean-field level in $\bar{H}^{ce}_{\rm MF}$. Finally, an effective chemical potential $\mu_{\rm emb}$ is added to preserve the total number of electrons in the fragment. 
All together, the cluster effective Hamiltonian reads 
  \begin{equation}
    \label{eq:cluster_Ham}
    \bar{H}^c_{\rm eff} =  \bar{H}^c + \bar{H}^{ce}_{\rm MF}+ \mu_{\rm emb}\sum_{i \in {\rm bath}} \tilde{c}_i^{\dagger}\tilde{c}_i,
  \end{equation}
  and 
  \begin{align}\label{eq:mf_ce}
    \bar{H}^{ce}_{\rm MF}[\gamma^\sigma] = & \sum_{i,j\in {\rm cluster} \atop \sigma }\tilde{c}^\dagger_{i\sigma}\tilde{c}_{j\sigma}\sum_{(kl)}\tilde{U}_{ijkl}\sum_{mn}\mathbf{R}^{\bar{\sigma}\dagger}_{km}\mathbf{R}^{\bar{\sigma}}_{nl}\gamma^{\bar{\sigma}}_{kl} \nonumber  \\
                                       & +{\rm h.c.} ,
\end{align}
where the notation $(kl)$ refers to pairs of orbitals in $\mathbf{R}^\sigma$ representation such that at least $k$ or $l$ belongs to the environment. Likely influenced by the non-interacting character of the bath in  DMFT, the non-locals interaction integrals $\tilde{U}_{ijkl}$ that naturally arise in the bath have been initially neglected in DMET, leading to a Anderson impurity model in the cluster~\cite{knizia_density_2012}. This approximation is called non-interacting bath (NIB) in contrast to the interacting bath (IB) version of DMET where $\tilde{U}_{ijkl}$ are explicitly taken into account~\cite{wouters_practical_2016}. The similar distinction between NIB an IB can also be considered using the Block-Householder transformation to construct the orbitals of the bath \cite{sekaran_householder_2021,marecat2023unitary}.


Interestingly for non-interacting Hamiltonian ($U = 0$), for which the associated  1RDM is idempotent ($\gamma^{\sigma 2} = \gamma^\sigma$), the transformation $\mathbf{R}^\sigma$ leads to a perfect decoupling of 1RDMs of the clusters (Eq.~(\ref{eq:uncoupl}) being fulfilled $\forall i \in $ the cluster instead of the fragment), respectively. It corresponds to an exact factorization of the underlying associated wave-function $|\Psi \rangle= |\phi \rangle$, where $|\phi \rangle$ refers to a single-Slater determinant. The factorization of the wave-function gives
\begin{equation}
  \mathbf{R}|\phi \rangle = \hat{\mathcal{A}}|\tilde{\phi}^c \rangle |\tilde{\phi}^e \rangle,
\end{equation}
where $ |\tilde{\phi}^c \rangle$ ($|\tilde{\phi}^e \rangle$) is an anti-symmetrized product of orbitals that belong solely to the cluster (environment). Consequently, in this specific case, the projected cluster Hamiltonian proposed in equation~(\ref{eq:cluster_Ham}) can be used to extract {\it exact} local properties of the impurity site, where the ground-state wave function $|\bar{\phi} \rangle$ is equal to $|\tilde{\phi}^c \rangle$~\cite{sekaran_householder_2021}.

In a more general context, particularly with interacting cases, the effective cluster Hamiltonian, as defined in equation~(\ref{eq:cluster_Ham}), can serve as a useful approximation for local properties of the fragment. This Hamiltonian is \textit{de facto} a functional of the 1RDM via the definition of $\mathbf{R}^\sigma[\gamma^\sigma]$. 
Several studies have developed self-consistent schemes predicated on the local cluster 1RDM calculated with equation~(\ref{eq:cluster_Ham}), as seen in DMET~\cite{knizia_density_2012} and more recent divide and conquer algorithms~\cite{marecat2023unitary}. 
Regardless of the self-consistent matching or conquer strategy employed, the efficiency of the method hinges crucially on the ability of the effective cluster Hamiltonian ~(\ref{eq:cluster_Ham}) to locally mimic the full Hamiltonian on the fragment. On this subject, there is no unambiguously definitive approach to determine what exactly constitutes "mimicry". To address this challenge, we propose in the following sections to generalize equation~(\ref{eq:V_Blk_House}), introducing variational parameters to explore different flavours of embedding.
\subsection{Quantum bath from a versatile unitary transformation framework}\label{sec:vur}

\begin{figure}
    \centering
    \resizebox{\columnwidth}{!}{
	\includegraphics[scale=0.20]{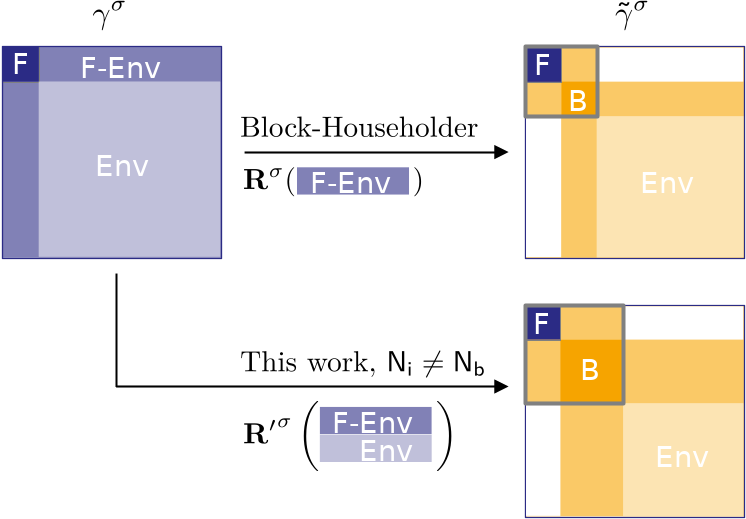}
    }
    \caption{Schematic representation of the Block-Householder transformation and the generalization, both functionals of the 1RDM. Small dark blue squares correspond to the fragment 1RDM in the system, identical as in the cluster, and orange squares correspond to bath 1RDM in the cluster. The gray squares correspond to the cluster.}
    \label{fig:flexible}
  \end{figure}

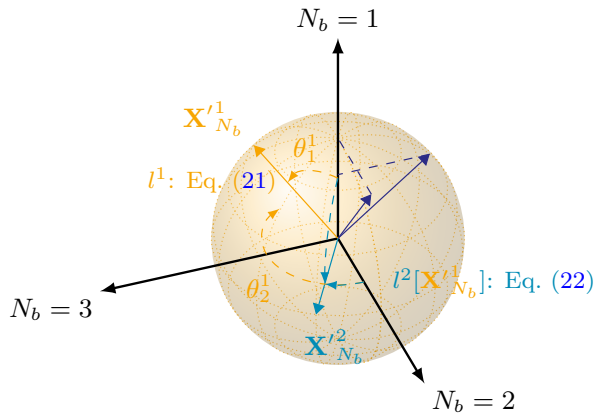
\begin{figure}
\centering
\resizebox{\columnwidth}{!}{
    
\tdplotsetmaincoords{90+37.6339922}{-90+20}
\begin{tikzpicture}[scale=3,tdplot_main_coords]
  
  \def\rvec{0.5}
  \def\r{\rvec}
  \def\thetavec{60}
  \def\phivec{-90+60}
    \def\thetablockA{50.44820236}
  \def\phiblockA{270}
    \def\thetablockB{37.6339922}
  \def\phiblockB{0.}
  \coordinate (O) at (0,0,0);
  \coordinate (gA) at (0,0,0.318318827);
  \coordinate (gB) at (0,0,0.5);
  \tdplotsetcoord {S}{0.5}{\thetablockA}{\phiblockA}
  \tdplotsetcoord {T}{0.63140567}{\thetablockB}{\phiblockB}
   \tdplotsetcoord{P}{\rvec-0.4}{\thetavec}{\phivec}
    \tdplotsetcoord{Q}{\rvec}{\thetavec}{\phivec+180}
    \tdplotsetcoord{R}{\rvec}{90}{30}
      \foreach \phi in {0, 30, ..., 180} {
    \draw [ICGMorange,densely dotted, domain=0:360, smooth, variable=\theta,opacity=0.6]
      plot ({\r*sin(\phi)*cos(\theta)}, {\r*sin(\phi)*sin(\theta)}, {\r*cos(\phi)});
  }
  \foreach \theta in {0, 30, ..., 360} {
    \draw [ICGMorange,densely dotted, domain=0:180, smooth, variable=\phi,opacity=0.6]
      plot ({\r*sin(\phi)*cos(\theta)}, {\r*sin(\phi)*sin(\theta)}, {\r*cos(\phi)});
  }
    \tdplotsetrotatedcoords{20}{90-\thetablockB}{0};
    \shade [ball color=ICGMorange,thin,tdplot_rotated_coords,opacity=0.25,draw=none] (0,0,0) circle (\rvec) ;
   

  \draw[thick,->] (0,0,0) -- (1,0,0) node[below right=-1]{$N_b=2$};
  \draw[thick,->] (0,0,0) -- (0,1,0) node[below left=-1]{$N_b=3$};
  \draw[thick,->] (0,0,0) -- (0,0,1) node[above=-1]{$N_b=1$};
  
  \draw[dashed,ICGMmarine] (S)  -- (gA) node[above left=-2] {};
  \draw[dashed,ICGMmarine] (T)  -- (gB) node[above left=-2] {};
      \draw[arrows={-Triangle},ICGMmarine] (O)  -- (S) node[above left=-2] {};
    \draw[arrows={-Triangle},ICGMmarine] (O)  -- (T) node[above left=-2] {};
   \draw[arrows={-Triangle},ICGMorange] (O)  -- (Q) node[above left] {$\mathbf{X'}^1_{N_b}$};
      \shade[ICGMorange] (O)  -- (Q) node[below=12, left=-11] { \footnotesize $l^1$: \Eq{eq:norm}};
 \draw[arrows={-Triangle},ICGMblue] (O)  -- (R) node[above left=-20] {  $\mathbf{X'}^2_{N_b}$};
       \shade[vector,ICGMblue] (O)  -- (R) node[right=60,above=2] { \footnotesize $l^2$[\textcolor{ICGMorange}{$\mathbf{X'}^1_{N_b}$}]: \Eq{eq:scalprod}};
  
  \tdplotdrawarc[dashed,->,color=ICGMorange]{(O)}{0.3}{0}{-\phivec}
    {anchor=north}{\hspace{-5mm} $\theta^1_2$}
  \tdplotsetthetaplanecoords{\phivec}
  \tdplotdrawarc[dashed,->,color=ICGMorange,tdplot_rotated_coords]{(0,0,0)}{0.3}{0}{-\thetavec}
    {anchor=south west}{\hspace{-3mm}$\theta^1_1$}

  \tdplotdrawarc[dashed,->,color=ICGMblue]{(O)}{0.3}{0}{30}
    {anchor=north}{\hspace{-5mm} }
  \tdplotsetthetaplanecoords{30}
  \tdplotdrawarc[dashed,->,color=ICGMblue,tdplot_rotated_coords]{(0,0,0)}{0.3}{0}{90}
    {anchor=south west,right=15,below=1}{\hspace{-3mm}}
  
\end{tikzpicture}
}
    \caption{Schematic representation of vectors $\mathbf{X'}^j_{N_b}$ (light blue and orange vectors) in spherical representation for $N_i=1$, $N_b=3$ and $r=2$. Dark blue vectors correspond to exact Block-Householder vectors from a density- matrix of a non-interacting 1D-Hubbard model. The length $l^1$ is fixed according to equation~(\ref{eq:norm}), while lengths $l^j$, $r \geq j > 1$ depends on the vector $\mathbf{X'}^1_{N_b}$ following the equation~(\ref{eq:scalprod}). All angles parameters $\{ \theta^j_i\}$, $1 \leq i \leq r-N_i$ are free. For example, all vectors belonging to the orange sphere are available as a choice of $\mathbf{X'}^1_{N_b}$. The number of free parameters is equal to $r\times(N_b-N_i)=4$ (orange and light blue dashed curves).}
    \label{fig:spher_rep}
  \end{figure}

Following the philosophy of exact diagonalization solver in DMFT~\cite{caffarel_exact_1994}, we would like to control the number of bath orbitals, independently of the number of impurity orbitals in the fragment, in order to systematically have a better description of the interactions of the fragment with the environment. As shown schematically in \Fig{fig:flexible}, the Block-Householder transformation leads the fragment (dark blue square) unchanged in the cluster (gray square), and interacts solely with the bath (orange square), where the number of bath orbitals is the same as in the fragment. In the following, we give a general and flexible definition of the unitary matrix $\mathbf{R'}^\sigma[\gamma^\sigma]$ to obtain an optimized cluster, composed of $N_i$ impurities coupled to $N_b\geq N_i$ bath orbitals.
In what follows, spin index $\sigma$  is omitted for clarity.
Similarly as the Block-Householder transformation, we use the auxiliary matrix $\mathbf{V'}$ with \\
\begin{eqnarray}\label{eq:transformation}
\mathbf{R'}^\sigma = Id. - 2\mathbf{V'}\left(\mathbf{V'}^T\mathbf{V'} \right)^{-1}\mathbf{V'}^T.
\end{eqnarray} 
The auxiliary matrix $\mathbf{V'} \in \mathbb{R}_{Nr}$, where $N_i \leq r \leq N$ refers to the rank of the auxiliary matrix, is constructed as following\\
\begin{eqnarray} \label{eq:V}
\mathbf{V'} = \left(
\begin{array}{l}
\mathbf{0}^r_{N_i}\\
\gamma^r_{N_b} - \mathbf{X'}^r_{N_b} \\
\gamma^r_{N_e}
\end{array}
\right),
\end{eqnarray}
and indexes $N_i$, $N_b$, $N_e$ refer to the number of impurity orbitals in the fragment, the number of bath orbitals and the number of orbitals in the environment, respectively.
$\gamma^r_{N_b}$ ($\gamma^r_{N_e}$) corresponds to the $r$ first columns of the spin 1RDM $\gamma^\sigma$ and the $N_b$ lines ($N_e$) after the $N_i$-th ($N_i+N_b$-th) first one.
At this point, condition (\ref{eq:iden}) (preservation of the fragment in the cluster) is already satisfied.
The matrix $\mathbf{X'}^r_{N_b}$ must be determined in order to preserve condition (\ref{eq:uncoupl}), {\it i.e.} it must satisfy the following equation,\\
\begin{eqnarray}\label{eq:build_X}
(\mathbf{X'}^r_{N_b} - \gamma^r_{N_b})^T(\mathbf{X'}^{N_i}_{N_b} + \gamma^{N_i}_{N_b}) = \gamma^{r^T}_{N_e}\gamma^{N_i}_{N_e} ,
\end{eqnarray}
where the superscript $N_i$ refers to the $N_i$ first column of the matrix.
In pratice, the nonlinear equation~(\ref{eq:build_X}) can be solved numerically.
In the following, we discuss only the single impurity case $N_i=1$, where equation~(\ref{eq:build_X}) can be solved analytically.
In this case, the matrix $\mathbf{X'}^{N_i}_{N_b}$ is reduced to a vector, and equation~(\ref{eq:build_X}) becomes \\
\begin{align}
&(\mathbf{X'}^1_{N_b} - \gamma^1_{N_b})^T(\mathbf{X'}^1_{N_b} + \gamma^1_{N_b}) = \gamma^{1^T}_{N_e}\gamma^1_{N_e}  \label{eq:non-lin}\\
&(\mathbf{X'}^j_{N_b} - \gamma^j_{N_b})^T(\mathbf{X'}^1_{N_b} + \gamma^1_{N_b}) = \gamma^{j^T}_{N_e}\gamma^1_{N_e}, \, \forall \, 1<j\leq r. \label{eq:lin}
\end{align}
Equation~(\ref{eq:non-lin}) is nonlinear and constrains the norm of the vector $\mathbf{X'}^1_{N_b}$. More precisely, this equation is fulfilled for any vectors $\mathbf{X'}^1_{N_b}$ satisfying\\
\begin{equation} \label{eq:norm}
|| \mathbf{X'}^{1}_{N_b}||=||\gamma^{1}_{N_b}|| + ||\gamma^{1}_{N_e}||,
\end{equation}
where $||\mathbf{v}||=\mathbf{v}^T\mathbf{v}$ refers to the norm of the vector $\mathbf{v}$.
Equation~(\ref{eq:lin}) exists only for $r>N_i$ and is linear for $\mathbf{X'}^j_{N_b}$, corresponding to a scalar-product preservation, and can be written as follows,\\
\begin{eqnarray}\label{eq:scalprod}
\mathbf{X'}^{j^T}_{N_b}\mathbf{X'}^1_{N_b} = \gamma^{j^T}_{N_b}(\mathbf{X'}^1_{N_b} + \gamma^1_{N_b}) + ||\gamma^{1}_{N_e}||.
\end{eqnarray}
We introduce a spherical representation which allows to express the vectors $\mathbf{X'}^j_{N_b}$ with lengths $l^j$ and a complete set of angles $\{\theta\}$. 
In this representation, length $l^1$ ($l^j$) is used to fulfill norm preservation~(\ref{eq:norm}) (scalar product~(\ref{eq:scalprod})) for $\mathbf{X'}^1_{N_b}$ ($\mathbf{X'}^j_{N_b}$) respectively. 
The set of angles $\{\theta\}$ are thus completely free and can take any value between $[-\pi,\pi[$. 
Consequently, we have a complete set of $\{\theta\}$ parameters to construct the corresponding set of different auxiliary matrices defined in equation~(\ref{eq:V}) satisfying the conditions~(\ref{eq:iden}) and~(\ref{eq:uncoupl}). The number of free parameters is equal to $r\times(N_b-N_i)$, with $r \geq N_i$.
In \Fig{fig:spher_rep}, we illustrate all vectors that can be obtain for $N_i=1$, $N_b=3$ and $r=2$.
Dark blue vectors correspond to the special Block-Householder solution, where $N_b=1$ corresponds to the second impurity in fragment. The rank two Block-Householder transformation preserves the identity over the second impurity. As a result, the projection of the first (second) Block-Householder vector over the second impurity $i=2$ axis gives $\gamma^\sigma_{12}$ ($\gamma^\sigma_{22}$) , as shown with the dashed dark blue line in \Fig{fig:spher_rep}. Beyond the special Block-Householder transformation, all vectors belonging to the orange sphere are solution of equation~(\ref{eq:norm}) such as presented with $\mathbf{X'}^1_{N_b}$ (orange vector) for example. The second vector (light blue vector) norm depends on its scalar product with $\mathbf{X'}^1_{N_b}$. Thus the norm (and the direction) is fixed using equation~(\ref{eq:scalprod}) and all other angles left are free.
More generally, when $r>1$, every vectors $\mathbf{X'}^j_{N_b}$ , $1<j\leq r$ are independent from each other and are correlated to $\mathbf{X'}^1_{N_b}$  solely.
In the case of $r=N_i$, only one solution is available (with 0 degrees of freedom) and leads to the unique Block-Householder transformation.

Note that by considering $r$, $N_b$ and $N'_b>N_b$, the auxiliary matrix $\mathbf{V'}[\mathbf{X'}^r_{N_b}]$ space is not included into $\mathbf{V'}[\mathbf{X'}^r_{N'_b}]$, {\it i.e.} any rank $r$ auxiliary matrices expressed using equation~(\ref{eq:V}) with $N_b$ bath cannot be expressed with an auxiliary matrix with a greater number of bath $N'_b$. 
Similarly, if we consider $N_b$, $r$ and $r'>r$, we get $\mathbf{V'}[\mathbf{X'}^r_{N_b}] \notin \mathbf{V'}[\mathbf{X'}^{r'}_{N_b}]$, which means that different ranks $r$ lead to a a specific definition of the cluster.

At this stage we have proposed a generic construction of unitary transformations following equations~(\ref{eq:transformation}) and~(\ref{eq:V}) that generalize the block-Householder construction. Indeed, given an arbitrary number of bath orbitals $N_b$ ($N_b\geq N_i$) and rank $r$ of matrix $\mathbf{V'}$ ($N_i \leq r \leq N$), we  show that we can construct many unitary transformations that fulfill conditions~(\ref{eq:iden}) and~(\ref{eq:uncoupl}) up to  $r\times(N_b-N_i)$ free parameters. In the next section we propose to use these additional parameters in order to add physically motivated criteria to design bath orbitals. 

\subsection{Optimization of free parameters}\label{sec:ofp}

The free parameters $\{\theta\}$ are variationaly optimized to adjust the bath orbitals using physical insights, unlike the systematic Block-Householder transformation.
In what follows, the transformation $\mathbf{R'}^\sigma$ is a functional of the spin 1RDM, but also a function of a full set of free parameters $\{\theta\}$ and is denoted as $\mathbf{R'}^\sigma[\gamma^\sigma](\{\theta\})$, or in a more compact notation $\mathbf{R'}^\sigma(\theta)$.
The transformation of $\gamma^\sigma$ with the transformation $\mathbf{R'}^\sigma(\theta)$ is denoted as $\tilde{\gamma}^\sigma$.

According to the construction of $\mathbf{R'}^\sigma(\theta)$ fulfilling constraint~(\ref{eq:uncoupl}), the fragment is disconnected from the environment at the one-body level. In the non-interacting case, we show analytically that the Block-Householder ($r=N_i$) disconnects bath orbitals from the environment.
However, this is not the case for the correlated 1RDM. 
Therefore, the set of variational parameters is used to minimize the value of the so-called buffer-zone $\Delta$, which gives a quantitative insight into the disentanglement of the environment cluster at the one body level and leads to the saddle-point equation\\
\begin{align}
&\Delta^2(\theta)= \sum_{b \in \rm bath}\sum_{e \in \rm env} \tilde{\gamma}^{\sigma^2}_{ce}, \label{eq:buffer_zone}\\
&\dfrac{\partial \Delta(\theta_i)}{\partial \theta_i} = 0 \quad \forall \theta_i \in \{\theta\}.\label{eq:buffer_zone_saddle}
\end{align} 
Following a similar philosophy, one could minimize the single-particle Von-Neumann entropy of the truncated 1RDM of the cluster.

From the medium to the strong correlated regime, the buffer-zone $\Delta$ is not able to give a quantitative value of the entanglement between the cluster and the environment, where two-body interactions dominate at this regime. 
In the following, we propose to minimize the square of the Hartree (i.e mean-field) contribution energy between the cluster and the environment \\
  \begin{eqnarray}\label{eq:Hartree}
E^2_{\rm H}(\theta) = \sum_{ c\in {\rm cluster} \atop \sigma}\left[\sum_{(jkl)}\left(\tilde{U}(\theta)_{cjkl}\tilde{\gamma}^\sigma_{cj} \tilde{\gamma}^\sigma_{kl} \right)^2 \right. \\
\left. + \sum_{ e\in {\rm env}}\left(\tilde{t}_{ce}\tilde{\gamma}^{\sigma}_{ce}\right)^2\right],\\
\label{eq:Hartree_saddle}
\dfrac{\partial E^2_{\rm H}(\theta_i) }{\partial \theta_i} = 0 \quad \forall \theta_i \in \{\theta\},
\end{eqnarray}
where $c$ ($e$) belongs to the cluster (environment), respectively, and the notation $(jkl)$ refers to pairs of orbitals in $\mathbf{R'}^\sigma(\theta)$ representation such that at least $j$, $k$ or $l$ belongs to the environment, and $\tilde{t}_{ce}$ ($\tilde{U}(\theta)_{cjkl}$) defined in equation~(\ref{eq:tce}) (equation~(\ref{eq:Uijkl})), respectively.
In a practical way, the evaluation of $E^2_{\rm H}$ using equation~(\ref{eq:Hartree}) is numerically more expensive than the evaluation of $\Delta$.

Finally, inspired by the DMET matching, we propose to design the transformation in order to enforce the matching between density matrix elements connected to the fragment in the transformation space, and in the cluster\\
\begin{equation}\label{eq:density_matching}
\underset{\theta}{\rm min}  \sum_{ij} \left(\tilde{\gamma}^\sigma_{ij} - \bar{\gamma}^{\sigma c}_{ij}\right)^2 \quad \forall i \in {\rm fragment, }\forall j \in {\rm cluster},
\end{equation}
with $ \bar{\gamma}^{\sigma c}$ refers as the ground-state spin $\sigma$ 1RDM of the cluster obtained by solving $\bar{H}^c_{eff}$ defined in equation~(\ref{eq:cluster_Ham}). 

\begin{figure}
     \centering
     \begin{subfigure}[b]{0.49\textwidth}
         \centering
        \includegraphics[scale=0.20]{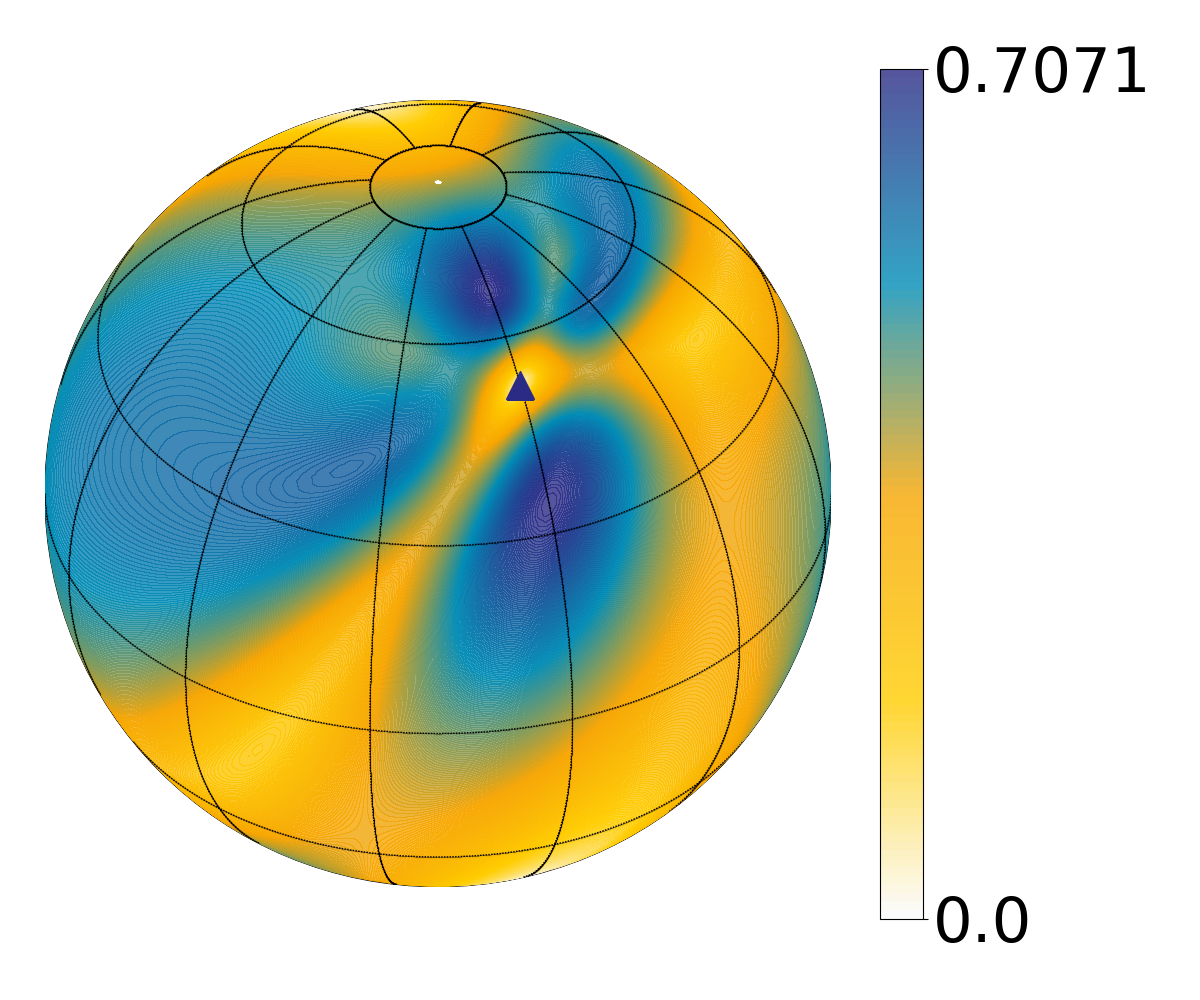}
         \caption{}
         \label{fig:buffer_zone}
     \end{subfigure}
     \begin{subfigure}[b]{0.49\textwidth}
         \centering
         \includegraphics[scale=0.20]{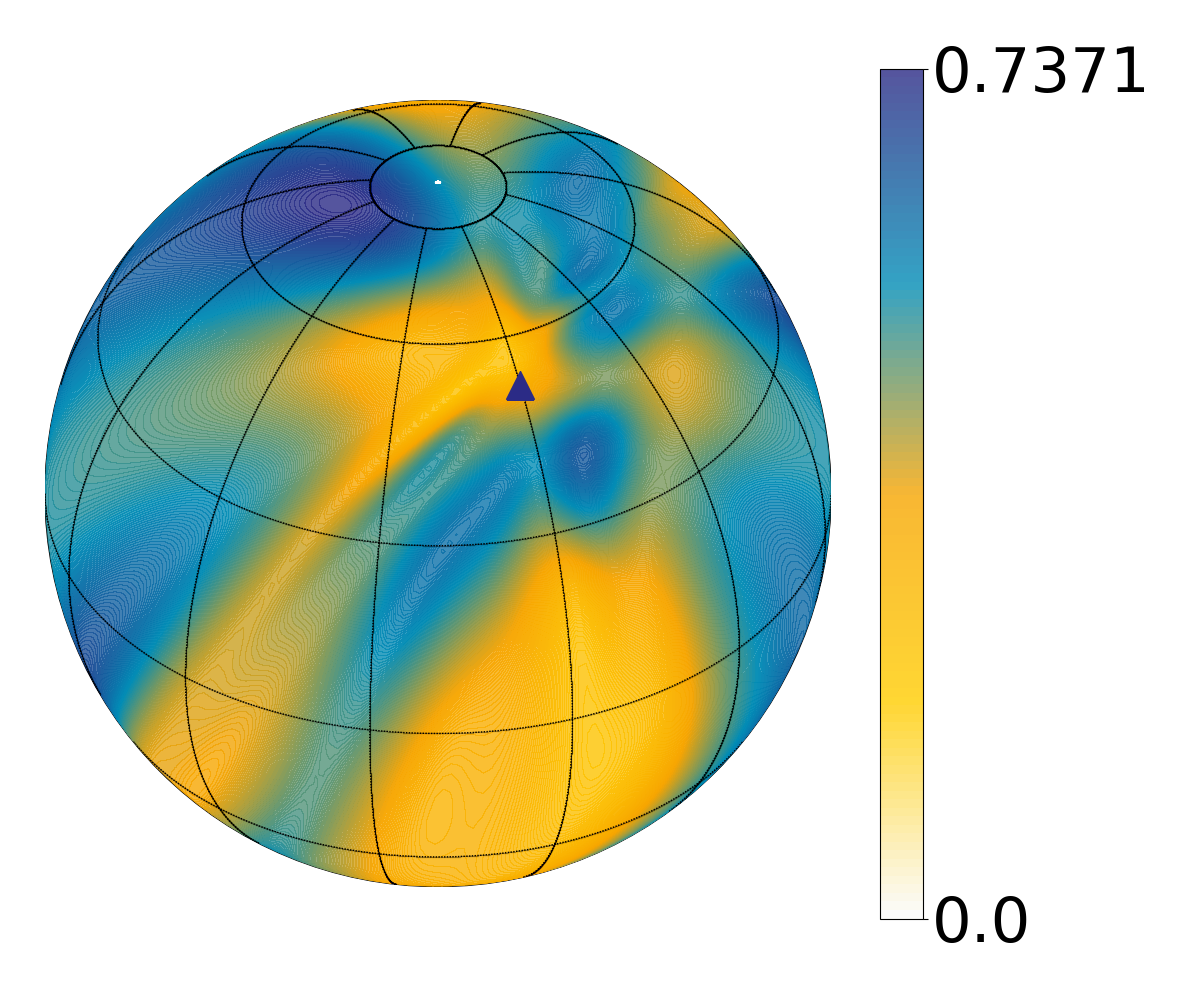}
         \caption{}
         \label{fig:Hartree}
     \end{subfigure}
     \begin{subfigure}[b]{0.49\textwidth}
     \vspace{0.2cm}
         \centering
         \includegraphics[scale=0.20]{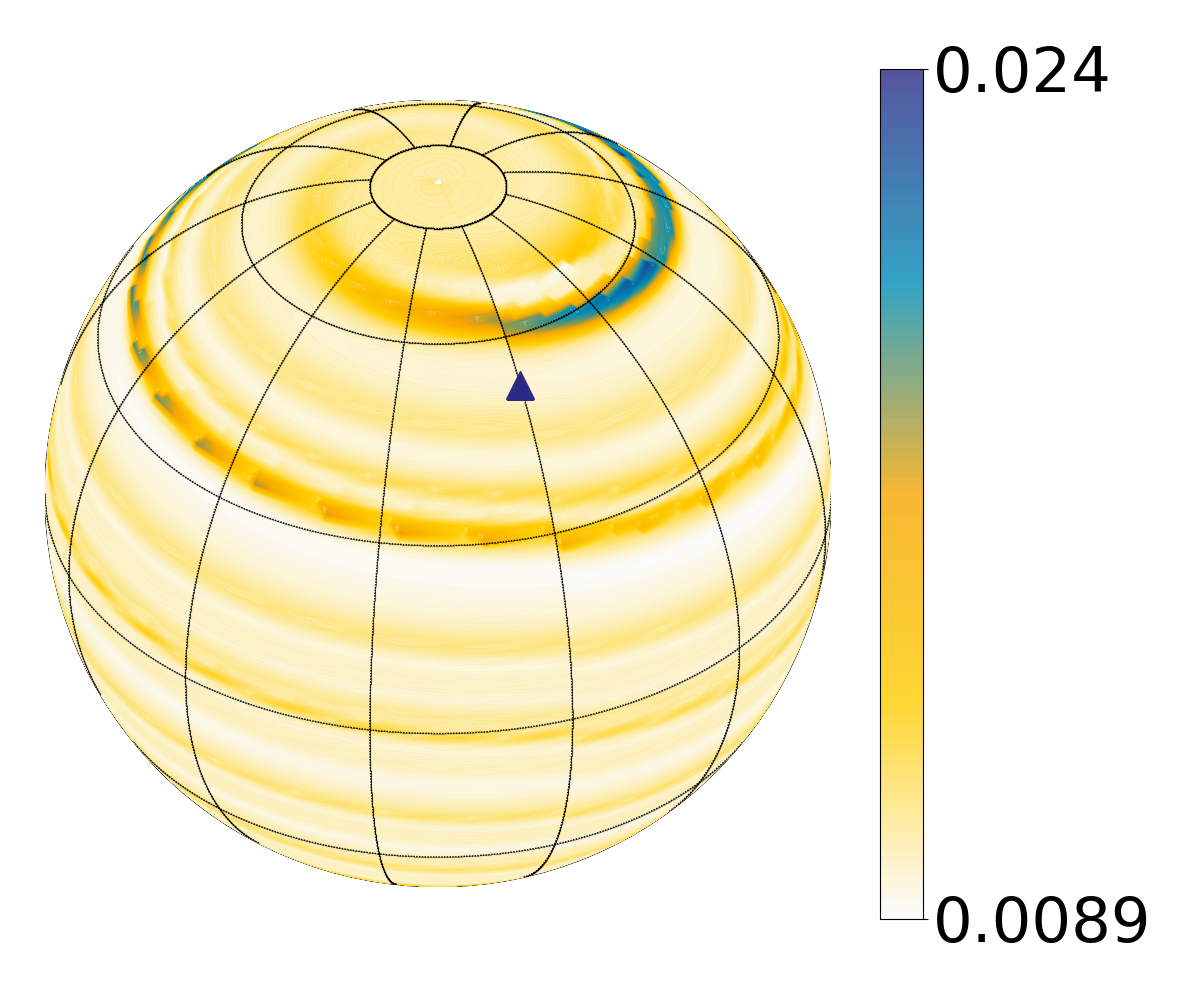}
         \caption{}
         \label{fig:density_matrix}
     \end{subfigure}
       \begin{subfigure}[b]{0.49\textwidth}
         \centering
         \caption*{Evaluation of the cost functions with respect to free parameters $\{\theta\}$ for $N_i=1$, $N_b=3$ and $r=2$ using a non-interacting one body reduced density matrix from a 1D Hubbard chain. The vector corresponding to the rank $r=1$ is fixed and corresponds to the Block-Householder vector (see \Fig{fig:spher_rep}), and the second is presented with the dark blue triangle. In this case, two angles are available, and are represented using spherical coordinates. 
In \textbf{(a)} we evaluate the buffer-zone $\Delta$ (see \Eq{eq:buffer_zone}). In \textbf{(b)}, we evaluate the Hartree contribution $E^2_{\rm H}$ (see \Eq{eq:Hartree}) for $U/t$ = 8, and in \textbf{(c)} we evaluate the density matrix matching (see \Eq{eq:density_matching}) for $U/t$ = 8.}
     \end{subfigure}
     \caption{}
        \label{fig:min_graphs}
      \end{figure}

As illustrated in \Fig{fig:min_graphs}, we test the cost functions proposed in equations~(\ref{eq:buffer_zone}),~(\ref{eq:Hartree}) and~(\ref{eq:density_matching}) with respect to two free parameters $\{\theta \}$ for $N_i=1$, $N_b=3$ and $r=2$.  A non-interacting one body reduced density matrix (idempotent) is used as a 1RDM test to obtain the transformation $\mathbf{R'}^\sigma(\theta)$. The first rank $r=1$ vector is fixed using Block-Householder solution. In that case, two free parameters are optimized, and cost functions can be represented in spherical coordinates. 
The minimization of the buffer zone $\Delta$ (\Fig{fig:buffer_zone}) strictly cancels the value of $\Delta$ and corresponds to the Block-Householder solution, as discussed in \Sec{sec:review}. This result is attributed to the fact that the trial density matrix is idempotent. However, this particular value is enclaved between regions of higher $\Delta$ values, which can make optimization difficult depending on the starting point of the numerical minimization. 
Subsequently, the minimization of the mean-field term between the cluster and the environment (\Fig{fig:Hartree}) also yields the same solution in this case. However, the landscape is distinct from the previously studied cost function.
Finally, the matching of the density matrices (\Fig{fig:density_matrix}) presents a very specific landscape. Indeed, such a landscape is numerically very challenging to explore in order to obtain the global minimum. Contrary to the cost functions studied previously, the Block-Householder solution (dark blue triangle) is not the global minimum. It appears that there exists a continuous set of minimums, which further complicates the exploration of the landscape.
Altogether, it highlights the non-trivial character of the resulting landscape that might display many local minima and quasi-flat regions. It results that the numerical optimization of the $\{\theta\}$ parameters might become challenging for large amount of parameters.

\section{Results and Discussion}
In this section, the various cost functions outlined in Section~\ref{sec:ofp} are evaluated against the homogeneous paramagnetic Hubbard model at half-filling, and compared to exact Bethe-Ansatz (BA) results~\cite{lieb_absence_1968,ogata1990bethe}.
The non-idempotent N-representable 1RDM space is spanned using the self-consistent protocol presented in~\cite{marecat2023unitary}.
The various results pertain to the case of a single impurity. The generalization to multiple impurities, discussed in Section \ref{sec:vur}, will not be covered in this paper.
We recall that the Block-Householder transformation corresponds to a particular case, where the set of parameters $\{\theta\}$ is set according to the equation~(\ref{eq:X}).
Additionally, the variational parameters are optimized using the numpy python library~\cite{2020SciPy-NMeth}, particularly with the L-BFGS-B method, which is similar to the conjugate gradient optimization method.
Finally, as explained in \cite{marecat2023unitary}, a damping parameter is added to avoid drastic changes of the 1RDM and convergence issues, meaning that a fraction of the previous 1RDM obtained is kept in the new 1RDM. In the results presented here we used a damping of $60\%$.

\begin{figure}
    \centering
    \resizebox{\columnwidth}{!}{
	\includegraphics[scale=0.20]{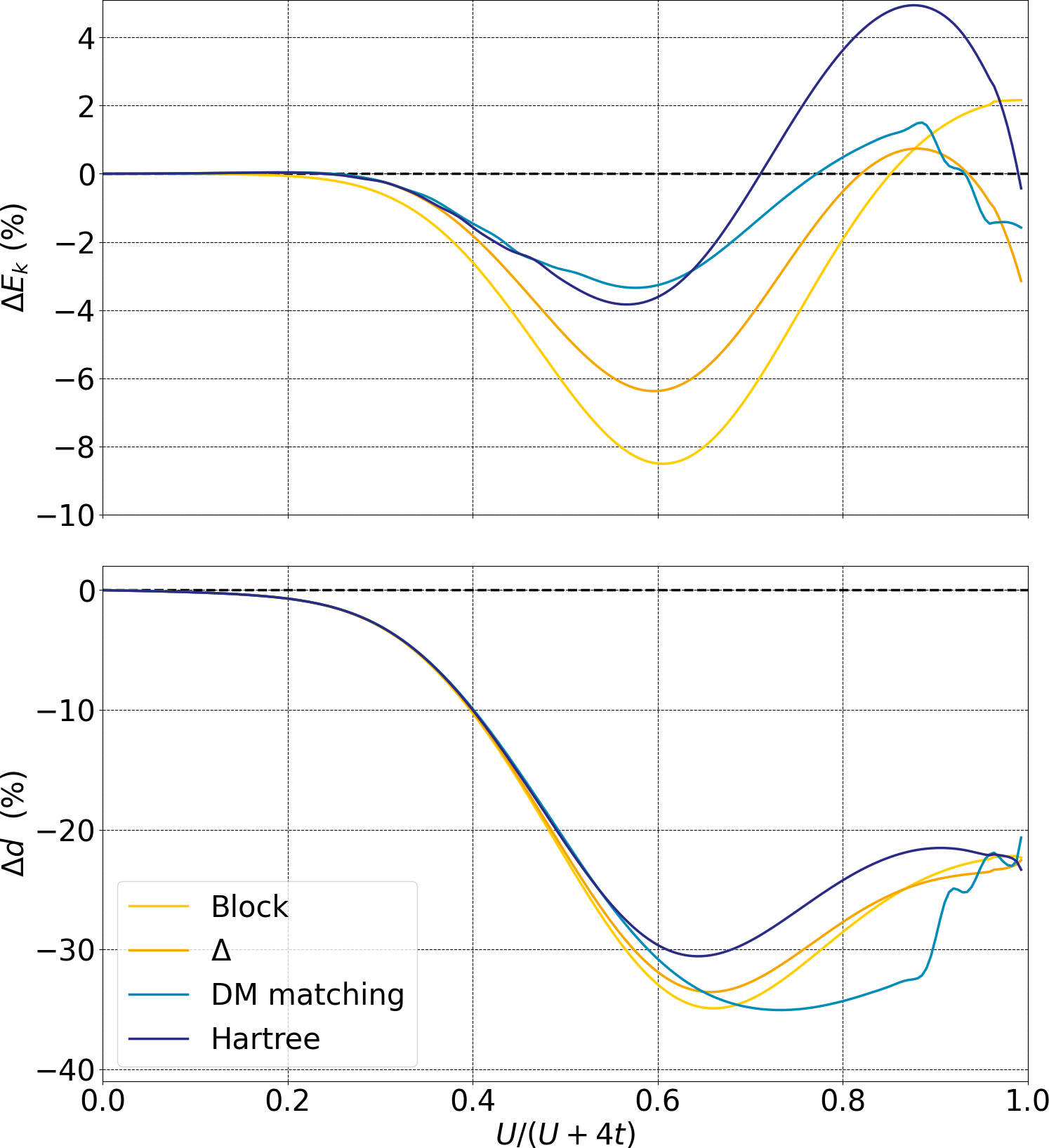}
    }
    \caption{Relative error for the kinetic energy $\Delta E_k$ (top panel) and per site double occupation $\Delta d$ (bottom panel) in percent with respect to correlation strength $U/(U+4t)$ for one impurity orbital in the fragment, three bath orbitals and a rank two vector. Colored lines correspond to different cost functions. Black dashed line corresponds to Bethe Ansatz.}
    \label{fig:energie_comp_minim}
  \end{figure}

In \Fig{fig:energie_comp_minim}, we show
 the relative error of the kinetic energy $\Delta E_k = 100\times \left( E^{\rm BA}_K - E_K \right)/E^{\rm BA}_K$, where BA refers to the Bethe Ansatz solution, and the relative error of the double occupation $\Delta d = 100\times \left( d^{\rm BA} - d \right)/d^{\rm BA}$
as a function of the relative correlation strength $U/(U+4t)$, where $4t$ corresponds to the non-interacting band width. 
This figure presents the effect of the various cost functions outlined in \Sec{sec:ofp} (color-coded lines). 
In this case, the single impurity $N_i=1$ is embedded with three IB orbitals $N_b=3$ for a vector $\mathbf{V'}$ of rank $r=2$. Thus, there is a number of parameters ${\theta}$ equal to $2 \times (3-1)=4$ to optimize.
In the non-interacting limit $U/t \rightarrow 0$, the kinetic energy and the double occupation are correctly reproduced for all cost functions. In regards to the atomic limit $U/t \rightarrow \infty$, this is also the case with an asymptotic limit of $-8$ln$(2)t^2/U$ for the kinetic energy, and $4$ln$(2)t^2/U^2$ for the double occupation~\cite{lopez-sandoval_density-matrix_2002}.
For intermediate regimes $U/4t \simeq 1$, there is a strong competition between electronic delocalization which increases kinetic energy, and the electron-electron repulsion strength which penalises the number of double occupation. For state-of-the-art embedding methods, such as DMET \cite{knizia_density_2012} or the projected site-occupation embedding theory (PSOET) \cite{senjean_projected_2019}, describing accuratly this regime is very challenging. 
For weakly to intermediate correlated regimes ($U/(U+4t)<0.6$, $U/t <6$), the results from the minimization of constrain~(\ref{eq:Hartree}) (dark blue line) are consistently better than those obtained by minimizing the constrain~(\ref{eq:buffer_zone}) (orange line), which are in turn better than the Block-Householder solutions (yellow line). 
However, the numerical cost associated with the minimization of constrain~(\ref{eq:Hartree}) scale as $\mathcal{O}\left(N_c N^4_e\right)$,  which is significantly higher than the cost associated with constrain~(\ref{eq:buffer_zone}) scaling as  $\mathcal{O}\left(N_c N^2_e \right)$, where $N_e$ ($N_c$) represents the number of orbitals belonging to the environment of the cluster (cluster), respectively.
Concerning the one body reduced density matrix matching proposed in equation~(\ref{eq:density_matching}) (blue line), the results obtained are similar to the constrain~(\ref{eq:Hartree}) from low to middle correlated regime, and deviate for values of $U/(U+4t)\simeq 0.6$ for kinetic energy and double occupancy.
For strongly correlated regimes ($U/(U+4t)\geq 0.6$), the interaction energy $E_{\rm int}=U\times d$ dominates. Although the double occupancy is similar for the Block-Householder method and the buffer-zone minimization at this regime, the minimization of the mean-field term improves the results. The double occupancy is badly described by the density matrix matching and exhibits nonphysical numerical instabilities for values of $U/(U+4t)\simeq 1$.

\begin{figure}
    \centering
    \resizebox{\columnwidth}{!}{
	\includegraphics[scale=0.20]{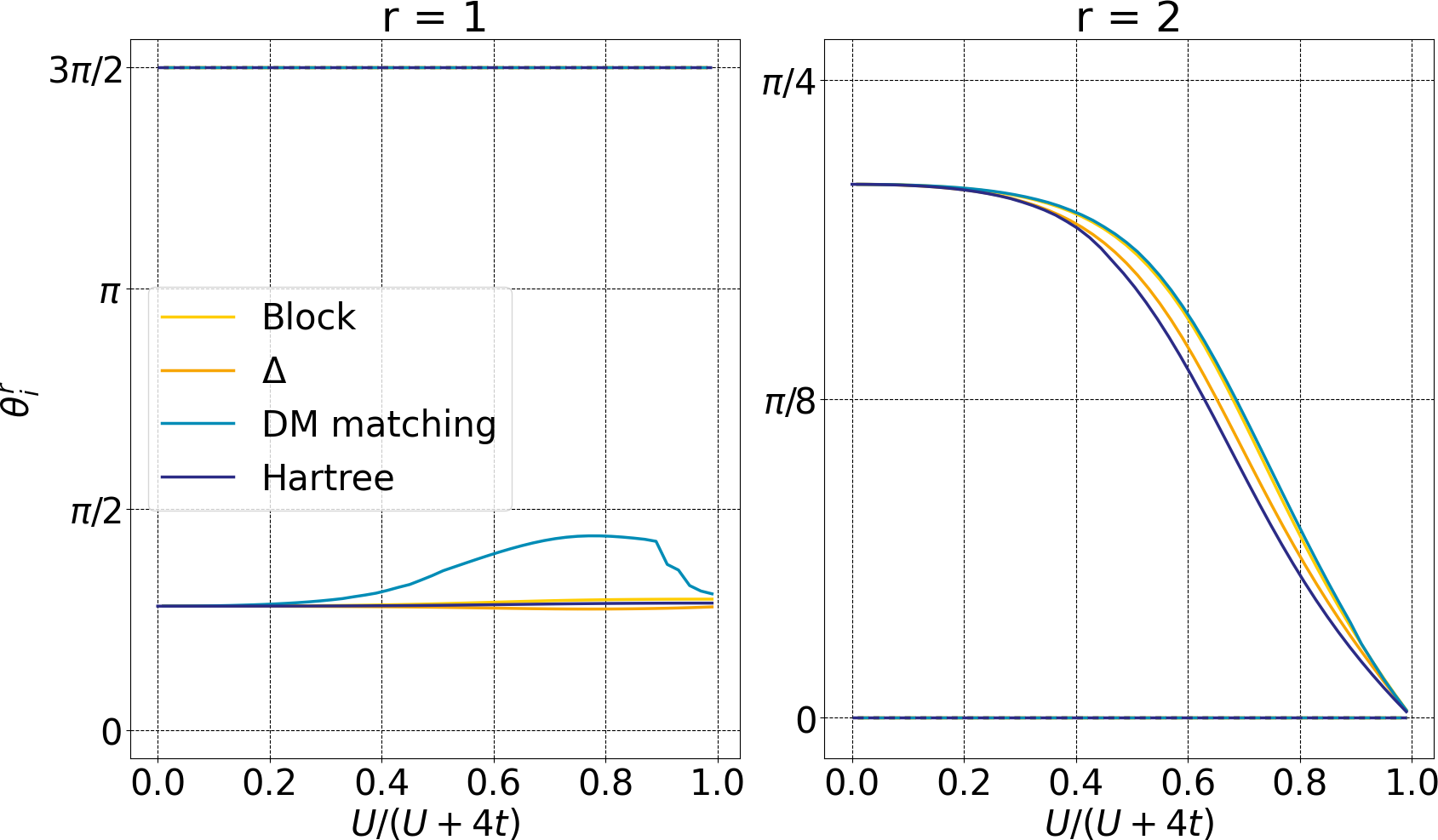}
    }
    \caption{Optimized free angle parameters with respect to the relative correlation strengh $U/(U+4t)$ for a rank $r=2$ and a three bath orbitals calculation, for different cost functions (colored dashed lines). The left panel corresponds to the first rank, while the right panel to the second. Solid lines correspond to the first parameter of each rank, and dashed lines to the second.}
    \label{fig:theta_urange}
  \end{figure}

In \Fig{fig:theta_urange}, we show the evolution of the different variational parameters ${\theta}$ for different cost functions as a function of relative correlation strengh $U/(U+4t)$ for a two rank $r$ and a three bath orbitals calculation. In the left panel, we show the two variational parameters associated with rank $r=1$, {\it i.e.}, the first column of the matrix $\mathbf{V'}$ in equation~(\ref{eq:V}), while in the right panel, the parameters are associated with rank $r=2$.
Solid lines correspond to the first angles for all ranks, and dotted lines to the second ones.
We found that all parameters except for the first of rank two (solid line in the right panel) do not change significantly with respect to $U/t$.
Interpreting this result is challenging. It might be due to specific symmetries of the system, such as translation invariance and electron-hole symmetry at half-filling.
Regarding the first parameters of rank two, they all follow the same trend, with the Hartree cost function slightly lower than the others.
In light of these results, it is conceivable to simplify the variational optimization of the different parameters by considering only a reduced set of parameters (in this case the first of the second rank) varying significantly in the process. This simplification could greatly improve the numerical optimization of the parameters, and therefore the numerical efficiency of the method in general.
  
\begin{figure}
    \centering
    \resizebox{\columnwidth}{!}{
	\includegraphics[scale=0.20]{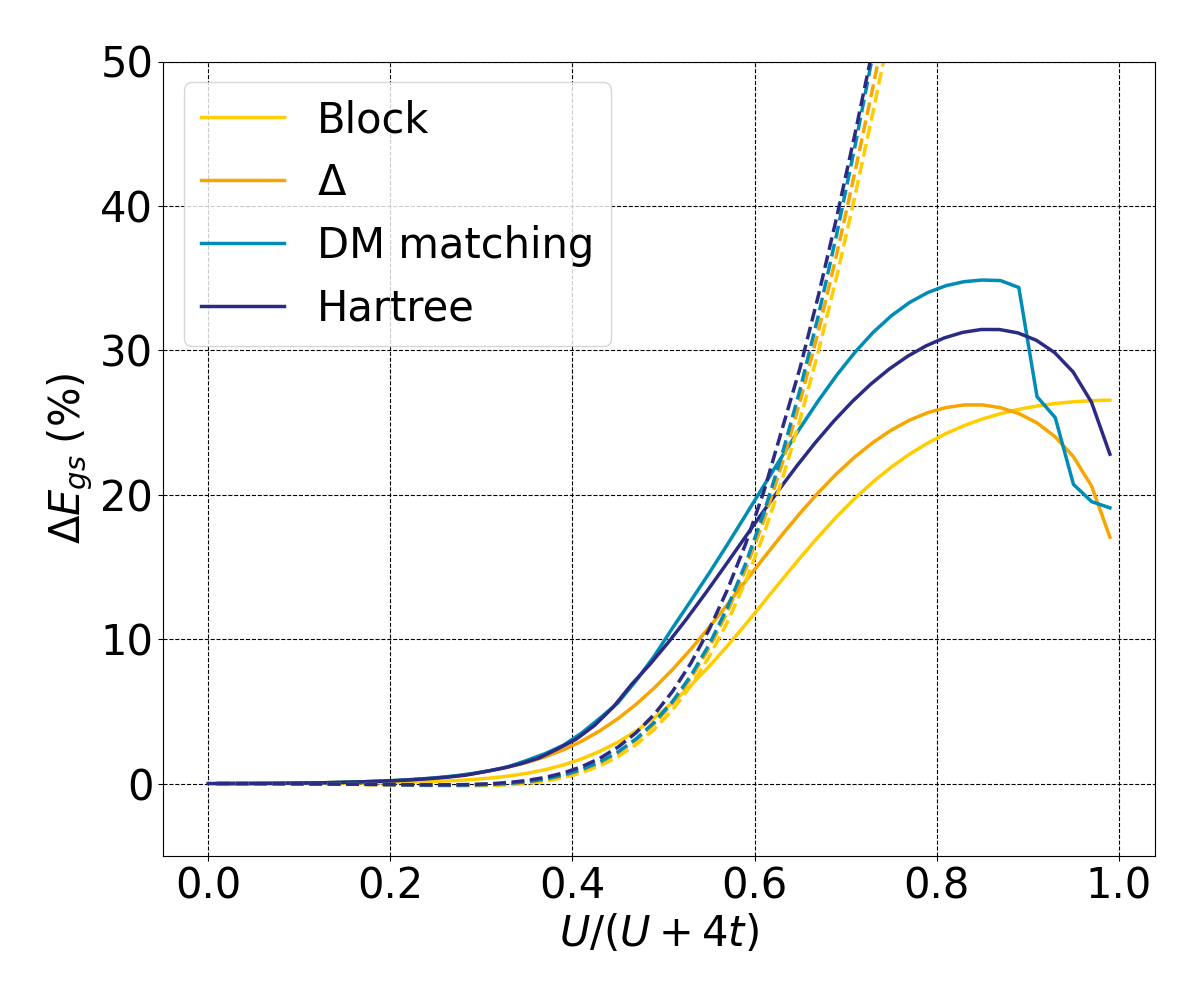}
    }
    \caption{Relative error of the ground state energy per site $\Delta E_{gs} = 100\times(1-E_{gs}/E_{BA})$ with respect to correlation strength $U/(U+4t)$ for one impurity and three bath orbitals and a rank two vector. Colored lines correspond to different cost functions for the IB case, and dashed lines for the NIB case. Black solid line corresponds to Bethe Ansatz.}
    \label{fig:energie_nib}
  \end{figure}

As originally presented in DMET~\cite{knizia_density_2012}, the effective Hamiltonian was an AIM, which implies that interaction terms within the bath are neglected, a scenario referred to as the NIB approximation. In this context, we present in \Fig{fig:energie_nib} the relative error of the ground-state energy per site $E_{gs} = E_k + Ud$, with respect to the correlation strength for a system with a single impurity in the fragment, three bath orbitals, and a rank-two vector. In the low correlation regime ($U/(U+4t)<0.3$), both NIB (represented by dashed colored lines) and IB (represented by full colored lines) yield similar results across all presented cost functions. However, for the intermediate to strong correlation regime ($U/(U+4t)>0.6$), the NIB approximation fails to provide an accurate description of the ground-state energy of the system. It should be noted that the Block IB (full yellow line) appears to offer the most accurate representation of the ground-state, contradicting the observations made in ~\Fig{fig:energie_comp_minim}. However, this apparent accuracy is misleading and arises from a larger error compensation between the kinetic and interaction energies. Additionally, we found the same error compensation for the NIB approximation, leading to inadequate results for both kinetic and interaction energies.
It is also noteworthy to optimize variational parameters to minimize the strength of Coulomb repulsion in the bath. This method could potentially close the gap between finite interacting baths in DMET or DaC approaches, and the infinite but non-interacting baths present in DMFT.
  
\begin{figure}
    \centering
    \resizebox{\columnwidth}{!}{
	\includegraphics[scale=0.20]{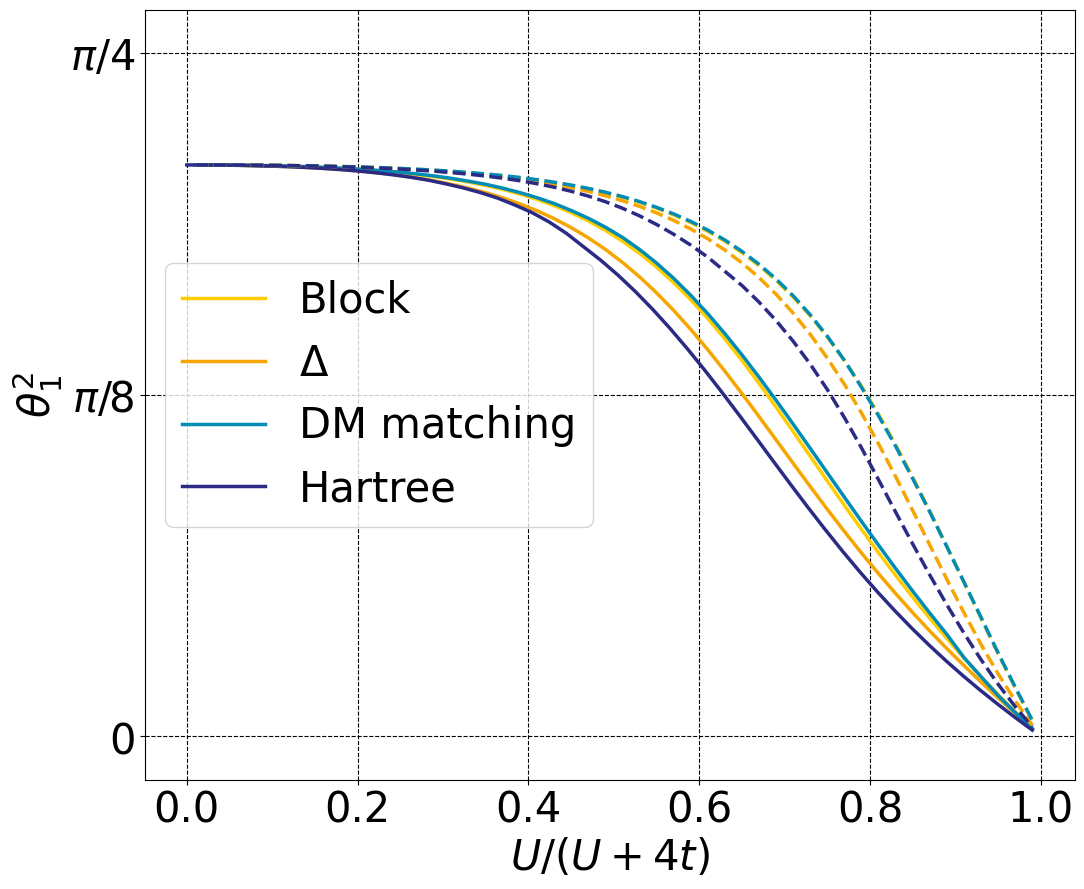}
    }
    \caption{Optimized free angle parameters with respect to the relative correlation strengh $U/(U+4t)$ for the first angle of the rank two vector of the set of parameters $\{\theta\}$, for different cost functions (colored lines). Solid lines correspond to the IB case, and dashed lines to the NIB case.}
    \label{fig:theta_urange_nib}
  \end{figure}

We present in \Fig{fig:theta_urange_nib} relevant optimized parameter (the first parameters of rank two, see results presented in \Fig{fig:theta_urange}) with respect to the relative correlation strengh for different cost functions (colored lines).
The two scenarios, NIB (dashed lines) and the IB (full lines), are presented.
For the NIB approximation, optimized parameters follow the same trend as the IB approximation, but are overestimated for all relative correlation strengh. As explained in \Fig{fig:energie_nib}, both NIB and IB yield similar results across all presented cost functions in the low correlated regime and equivalent parameters for both non-interacting ($U/t \rightarrow 0$) and atomic ($U/t \rightarrow \infty$) limits.
Interestingly, the interaction terms in the bath (IB), that emerge naturally with an embedding scheme based on a unitary transformation, are mandatory to obtain a correct description, in particular at large $U/t$. 

\begin{figure}
    \centering
    \resizebox{\columnwidth}{!}{
	\includegraphics[scale=0.20]{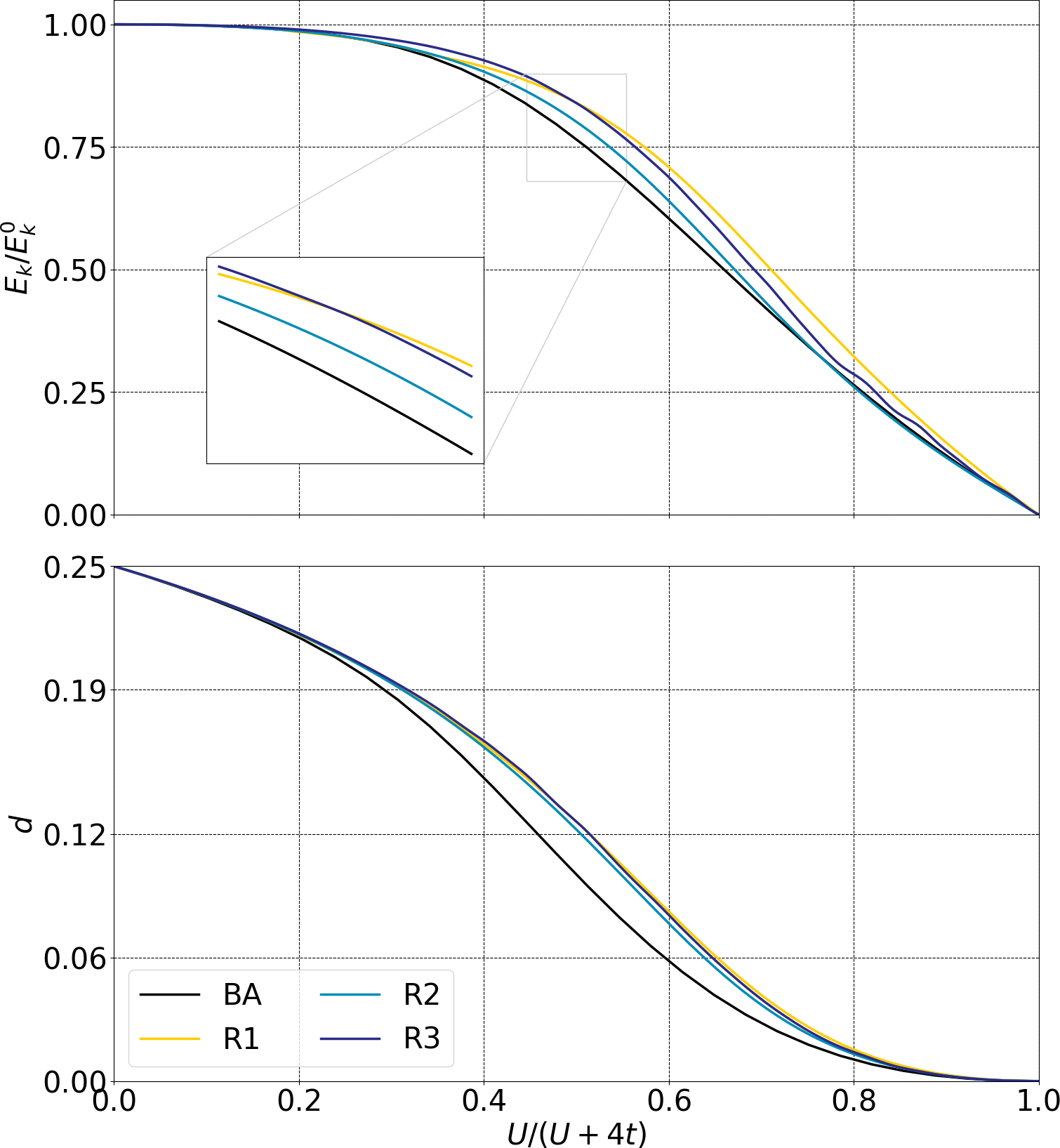}
    }
    \caption{Renormalized kinetic energy per site $E_k/E^0_k$ (top panel) and per site double occupation $d$ (bottom panel) with respect to correlation strength $U/(U+4t)$ for one impurity and three bath orbitals and the minimization of the buffer zone as a cost function. Colored lines correspond to the rank of the vector up to three. Black solid line correspond to Bethe Ansatz.}
    \label{fig:rank}
  \end{figure}

\begin{figure}
    \centering
    \resizebox{\columnwidth}{!}{
	\includegraphics[scale=0.20]{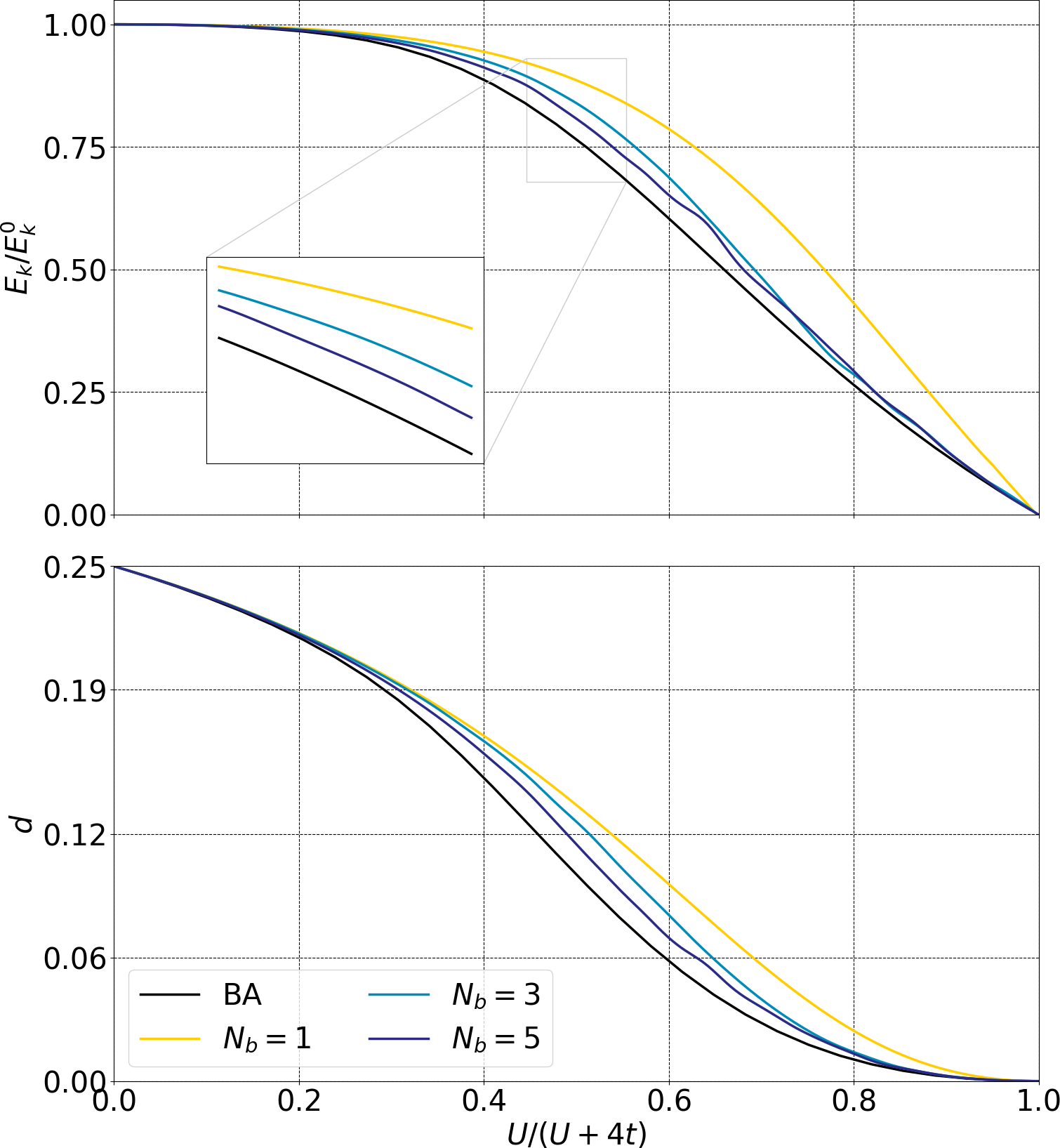}
    }
    \caption{Renormalized kinetic energy per site $E_k/E^0_k$ (top panel) and per site double occupation $d$ (bottom panel) with respect to correlation strength $U/(U+4t)$ for a rank three vector and the minimization of the buffer zone as a cost function. Colored lines correspond to the number of bath orbitals up to five. Black solid line corresponds to Bethe Ansatz.}
    \label{fig:bath}
  \end{figure}

 In the following, we focus on the IB case and the minimization of the buffer zone (see \Eq{eq:buffer_zone}) to explore the influence of the number of bath orbitals and the rank of the matrix  $\mathbf{V'}$ used to define the unitary transformation.
In \Fig{fig:rank}, we present the  per site kinetic energy scaled with the non-interacting kinetic energy per site $E^0_K=-4/\pi$ (upper panel) and the per site double occupation (lower panel) as a function of relative repulsion strength $U/(U+4t)$.
Results correspond to the cases $N_i=1$, $N_b=3$ and are given for different rank $r$ of the vector $\mathbf{V'}$, ranging from one up to three. For instance, the rank two (blue line) corresponds to the results previously shown in \Fig{fig:energie_comp_minim} (orange line).
The number of variational parameters is equal to $2$ for $r=1$, equal to $4$ for $r=2$, and equal to $6$ for $r=3$.
For a rank $1$ vector, we are in the special case where the rank equals the number of impurities $N_i$. In this case, the construction of vector $\mathbf{V'}$ needs to satisfy only the norm preservation in equation~(\ref{eq:norm}).
As demonstrated previously in Section \ref{sec:vur}, increasing the rank of vector $\mathbf{V'}$ does not systematically improve the solutions, as the accessible solution spaces are disjoint.
Indeed, we show that the rank $2$ results are the closest to the exact results for both kinetic energy and double occupation, and this applies to all correlation regimes $U/t$.
For strongly correlated regimes, rank $3$ slightly improves the results of rank $1$, but it exhibits numerical instabilities due to the optimization of a larger amount of variational parameters.
According to this figure, it is not necessary to increase the rank of vector $\mathbf{V'}$ in order to systematically improve the results.
Moreover, the best results are obtained for $r=2$, corresponding to the number of singular values in DMET~\cite{knizia_density_2012}, or the number of columns of the vector $\mathbf{V'}=\mathbf{V}$ for the Block-Householder method presented here~\cite{marecat2023unitary}.
In \Fig{fig:bath}, we present the scaled kinetic energy per site (upper panel) and the per site double occupation (lower panel) as a function of relative repulsion strength $U/(U+4t)$.
The results are shown for a rank $r=3$ vector $\mathbf{V'}$, for different number of orbitals in the bath, ranging from one up to five (colored lines). For spin symmetry reasons, we only consider cases where the number of orbitals in the cluster is even. The case with a single orbital in the bath (yellow line) is very particular, as there are no variational parameters to be optimized in this case. In the other cases, the number of variational parameters corresponds to $6$ for $N_b = 3$ orbitals, and $12$ for $N_b = 5$ orbitals. Importantly, increasing the number of orbitals in the bath systematically improves the kinetic energy and double occupation for all correlation regimes $U/t$. However, for strongly correlated regimes $U/t>>1$, we observe oscillations of the solutions for five bath orbitals. In the latter case, there are a large number of variational parameters, and their optimization is numerically challenging with the proposed iterative process.
\section{Conclusion}
In this study, we have thoroughly investigated the performance of various 1RDM based embedding methods to construct the orbitals of the bath. A particular emphasis is placed on the characteristics of the resulting reduced and effective Hamiltonian. Indeed this Hamiltonian is tasked with accurately reproducing the interactions between the fragment of the system and its environment within a downscaled cluster. 

While DMET employs the SVD of the fragment-environment 1RDM to define the effective Hamiltonian, we have demonstrated that the compact subspace is not the optimal setting for deriving the effective Hamiltonian. By generalizing the Block-Householder equations, we introduce a significant amount of additional flexible parameters, notably by adding bath orbitals that are nearly independent of the number of fragment orbitals or by exploring different transformation domains via rank augmentation.
To efficiently leverage these additional degrees of freedom, we have proposed cost functions that, in most cases, effectively disconnect the cluster, containing an integer number of electrons, from the environment. 

These cost functions were tested on the half-filled Hubbard model Hamiltonian, with a single impurity orbital in the fragment for which the equations are simplified. The results showed significant improvements over the Block-Householder outcomes.

Nevertheless, these improvements imply numerical optimization, which often proves challenging due to the complex landscape of cost functions. The complexity of these landscapes likely contributes to the fact that we can currently only achieve half-filled results. Moreover, this complexity occasionally makes it difficult to obtain continuous solutions for all relative correlation strengths, resulting in certain non-physical instabilities. Therefore, we encourage further research into the development of efficient cost functions that can derive an optimized effective Hamiltonian to describe the fragment, offering a smoother landscape than its counterparts.
A compelling challenge for future research would be to test the method on multiple-impurity fragments and, importantly, to derive linearized equations to define the unitary transformation more effectively.
\begin{acknowledgments}
The authors would like to thank the ANR (Grant No. ANR-19-CE29-0002 DESCARTES
project) for funding.
\end{acknowledgments}

%
\end{document}